\documentclass[aps,pre,superscriptaddress,showkeys,twocolumn,longbibliography]{revtex4-2}

\usepackage{amsmath}
\usepackage{graphicx}
\usepackage{float,physics}
\usepackage{siunitx}
\usepackage{xcolor}
\usepackage{epstopdf}
\definecolor{darkgreen}{rgb}{0,0.6,0.0}

\usepackage{latexsym}
\usepackage{amsmath}
\usepackage{amsfonts}
\usepackage{amssymb}
\usepackage{mathrsfs}
\usepackage{verbatim}
\usepackage{anysize}
\usepackage{soul}

\usepackage[utf8]{inputenc}
\usepackage{algorithmic}
\usepackage{algorithm}
\usepackage{enumerate}
\usepackage{amssymb, amsmath, amsthm}
\usepackage{graphicx}
\usepackage{lmodern,url}
\usepackage{graphicx}
\usepackage{wrapfig}
\usepackage{hyperref}

\floatname{algorithm}{Algoritmo}

\newcommand{\bo}{\raise-1mm\hbox{\Large$\Box$}}

\usepackage{multirow}

\usepackage{hyperref}
\hypersetup{
    colorlinks=true,
    linkcolor=black,    
    citecolor=blue,    
    urlcolor=blue       
}

\begin{document}
\title{Restoring the fluctuation-dissipation theorem in Kardar-Parisi-Zhang universality class through a new emergent fractal dimension}
\normalsize
\author{M\'arcio S. Gomes-Filho}
\affiliation{ICTP-South American Institute for Fundamental Research - Instituto de F\'isica Te\'orica da UNESP, Rua Dr.~Bento Teobaldo Ferraz 271, 01140-070 S\~ao Paulo, Brazil.}
\affiliation{Centro de Ciências Naturais e Humanas, Universidade Federal do ABC, 09210-580, Santo André, São Paulo, Brazil}
\author{Pablo de Castro}
\affiliation{ICTP-South American Institute for Fundamental Research - Instituto de F\'isica Te\'orica da UNESP, Rua Dr.~Bento Teobaldo Ferraz 271, 01140-070 S\~ao Paulo, Brazil.}
\author{Danilo B.\ Liarte}
\affiliation{ICTP-South American Institute for Fundamental Research - Instituto de F\'isica Te\'orica da UNESP, Rua Dr.~Bento Teobaldo Ferraz 271, 01140-070 S\~ao Paulo, Brazil.}
\author{Fernando A.\ Oliveira}
\email[]{faooliveira@gmail.com }
\affiliation{Instituto de F\'isica, Universidade de Bras\'ilia, Bras\'ilia-DF, Brazil}
\affiliation{Instituto de F\'isica, Universidade Federal Fluminense, Avenida Litor\^anea s/n, 24210-340, Niter\'oi, RJ, Brazil}
\begin{abstract}
The Kardar-Parisi-Zhang (KPZ) equation describes a wide range of growth-like phenomena, with applications in physics, chemistry and biology. There are three central questions in the study of KPZ growth: the determination of height probability distributions; the search for ever more precise universal growth exponents; and the apparent absence of a fluctuation-dissipation theorem (FDT) for spatial dimension $d>1$. Notably, 
these questions were answered exactly only for $1+1$ dimensions. {In this work, we propose a new FDT valid for the KPZ problem in $d+1$ dimensions. This is done by rearranging terms and identifying a new correlated noise which we argue to be characterized by a fractal dimension $d_n$}. We present relations between the KPZ exponents and two emergent fractal dimensions, namely $d_f$, of the rough interface, and $d_n$. Also, we simulate KPZ growth to obtain values for transient versions of the roughness exponent $\alpha$, the surface fractal dimension $d_f$ and, through our relations, the noise fractal dimension $d_n$. {Our results indicate that KPZ may have at least two fractal dimensions and that, within this proposal, a FDT is restored.} Finally, we provide new insights into the old question about the upper critical dimension of the KPZ universality class.
\end{abstract}
\maketitle
\section{Introduction}
Many major advances in physics have involved a clear understanding of the connections between physical laws and geometry. For instance, the classical mechanics revolution led by Galileo and Newton became possible with the development of calculus applied to Euclidean geometry. Similarly, in the realm of quantum mechanics, fundamental concepts such as symmetry and groups are linked to geometric principles. In general relativity, the connection between physics and geometry is so profound that one determines the other.  

However, Mandelbrot's fractal revolution in complex systems~\cite{Mandelbrot82} is somewhat incomplete. This incompleteness is related to the intricate nature of complex systems, which can span various spatial and temporal scales, often exhibiting diverse regimes of relaxation processes. The issue is that, in general, we do not know how to deal with fractal geometries exactly. In fact, exact fractal dimensions are known only for some deterministic objects with previously defined scaling rules. Even approximate numerical methods should be used carefully~\cite{Luis22,Luis23}.
For stochastic variables, such scaling rules are typically unknown and valid only statistically. Nevertheless, concepts of fractality continue to arise in  physics~\cite{Zaslavsky05}. In particular, fractal dimensions often emerge in the fundamental phenomenon of diffusion~\cite{Oliveira19}. Fractals also emerge in problems of growing surfaces, as discussed in this work.

In many physical systems, growth processes can occur as particles or aggregates of particles reach a surface through diffusion or some other form of deposition process, or even an injection beam. To investigate this growth, one tracks the height $h(\vec{x},t)$ of the growing surface, where $t$ is time, and $\vec{x}$ is the position in a space of dimension $d$. Since $h(\vec{x},t)$ typically exhibits scaling properties different from $\vec{x}$, we refer to $(h(\vec{x},t),\vec{x})$ as forming a $d+1$ dimensional space. Field equations have been proposed for the dynamics of $h(\vec{x},t)$, such as the Kardar-Parisi-Zhang (KPZ) equation~\cite{Kardar86}:
\begin{equation}
\label{KPZ}
\dfrac{\partial h(\vec{x},t)}{\partial t}=\nu \nabla^2 h(\vec{x},t) +\dfrac{\lambda}{2}[\vec{\nabla}h(\vec{x},t)]^2+ \eta(\vec{x},t).
\end{equation}
{The coefficient  $\nu$ is a surface tension parameter that controls a diffusive-like term associated with the so-called Laplacian smoothening mechanism. The term with $\lambda$ is nonlinear and related to the tilt mechanism (lateral growth).}
The  Gaussian white noise, $\eta(\vec{x},t)$,  has zero mean $\langle \eta(\vec{x},t) \rangle = 0$ and variance
\begin{equation}
\label{FDT}
\langle \eta(\vec{x},t) \eta(\vec{x'},t') \rangle  = 2D\delta^{(d)}(\vec{x}-\vec{x'})\delta(t-t'),
\end{equation}
{where $D$ controls the noise intensity~\cite{Edwards82,Kardar86} and  $\langle \cdots \rangle$ denotes an ensemble average. 
For $\lambda=0$, the Edwards-Wilkinson (EW) equation is recovered~\cite{Edwards82}. 
{The KPZ equation describes and connects a broad spectrum of significant stochastic growth-like processes in physics, chemistry, and biology, spanning from classical to quantum systems (see discussions and references in~\cite{GomesFilho21b,Barabasi95}). From time to time, a new system is discovered to belong to the KPZ universality class.}

A large number of such growth-like phenomena~\cite{Barabasi95,Merikoski03,Doussal16,Orrillo17,Ojeda00,Chen16,rojas2023wetting} can be understood by defining a few physical quantities such as the average height $\langle h \rangle$ and  the roughness or surface width
\begin{equation}
\label{W2}
 w(L,t)^2= \langle h^2(t)\rangle -\langle h(t) \rangle^2,
\end{equation}
where $L$ is the linear sample size. We are interested in physical systems in which the roughness grows with time and then saturates at a maximum value $w_s$ \cite{Barabasi95}:
\begin{equation}
\label{Sc1}
w(L,t)\approx
\begin{cases}
 ct^\beta , &\text{ if~~ } t \ll t_\times\\
 w_s, &\text{ if~~ } t \gg t_\times,\\
\end{cases}
\end{equation}
with $w_s \sim L^\alpha$ and $t_\times \sim L^z$, where $t_\times$ is a crossover time.
The critical exponents $z$, $\alpha$ and $\beta$ satisfy the scaling relation~\cite{Family85,Rodrigues24} 
\begin{equation}
\label{z}
z=\frac{\alpha}{\beta}.
\end{equation}
Also, the one-loop renormalization group approach preserves Galilean invariance, which results in to~\cite{Kardar86}
\begin{equation}
\label{GI}
\alpha+z=2,
\end{equation}
and therefore there is only one independent exponent.

\section{The fluctuation-dissipation theorem}
Our starting point is to try to understand the fluctuation-dissipation theorem (FDT) in KPZ growth systems. Since there is a long history of {violation} of the FDT in some complex systems such as structural glasses~\cite{Grigera99,Ricci-Tersenghi00,Barrat98,Bellon02}, proteins~\cite{Hayashi07}, 
mesoscopic radioactive heat transfer ~\cite{Perez-Madrid09} and ballistic diffusion~\cite{Costa03,Costa06,Lapas07,Lapas08,villa2020run}, it has been suggested that for KPZ, the FDT should always fail at dimension $d>1$~\cite{Kardar86,Rodriguez19,GomesFilho21,GomesFilho21b,Anjos21} (for a review, see \cite{gomesfilho23}). 

More recently, we demonstrated the existence of a FDT for KPZ growth in 1+1 dimensions~\cite{GomesFilho21}, leading us to find the corresponding KPZ exponents for $2+1$ dimensions analytically~\cite{Anjos21}. We explored the idea that the fractal dimension of the surface, denoted $d_f$, is connected to the KPZ exponents at the saturation of the growth process. 
This connection allowed us to derive precise exponents compared to numerical and experimental results, particularly for 2+1 dimensions~\cite{GomesFilho21b}. 
 Here, we discuss a new emergent fractal dimension directed associated to the noise of the process, denoted as $d_n$, which emerges from the dynamics, and how both fractal dimensions are related to the critical exponents.

This apparent violation of the FDT at higher dimensions motivates us to look more carefully into the KPZ equation. First, note that since $[\vec{\nabla}h(\vec{x},t)]^2 > 0$, the nonlinear term always carries the sign of $\lambda$, contrasting with the Laplacian and noise contributions, which in turn fluctuate between positive and negative.
Note as well  that the average growth velocity $v_g$ is given by~\cite{Barabasi95}
\begin{equation}
    v_g= \frac{\lambda}{2}\langle [\vec{\nabla}h(\vec{x},t)]^2 \rangle. 
\end{equation}
Our time is measured in deposition layer units in such a way that $v_g$ is constant.
 Thus, we rewrite Eq.\ (\ref{KPZ}) as
\begin{equation}
\label{KPZm}
\dfrac{\partial h(\vec{x},t)}{\partial t}=\nu \nabla^2 h(\vec{x},t) +v_g+ \phi(\vec{x},t),
\end{equation} 
which results in an Edwards-Wilkinson equation \cite{vvedensky2003edwards} with constant velocity and effective noise
\begin{equation}
\label{phi}
\phi(\vec{x},t)=\eta(\vec{x},t)+\psi(\vec{x},t)
\end{equation}
where
\begin{equation}
    \psi(\vec{x},t)= \frac{\lambda}{2} [\vec{\nabla}h(\vec{x},t)]^2-v_g. 
\end{equation}
$\psi(\vec{x},t)$ is just the fluctuation of the nonlinear term. Observe that the original noise $\eta(\vec{x},t)$ is uncorrelated in time and space as presented in Eq.\ (\ref{FDT}), whereas $\psi(\vec{x},t)$ is a noise strongly correlated in space with first neighbors, which can be concluded from its definition. Note that, by construction, $\langle \phi(\vec{x},t) \rangle =0$.

We note that, since the growth process usually starts with a flat  surface $h(\vec{x},t=0)=0$, the initial noise is just  $\phi(\vec{x},t=0)=\eta(\vec{x},t=0)$ and the first state of the growth is just a random walk. It is followed by a correlation such that $w(t) \propto t^\beta$, where distinctions between Edwards-Wilkinson and KPZ appear.
The distribution of heights $P(h)$, which has been obtained exactly only for $1+1$ dimensions and shows universal behavior~\cite{Doussal16,Calabrese10,Amir11,Prahofer00,Sasamoto10}, will dynamically affect the noise $\phi(\vec{x},t)$  and the  roughness of the interface.

\subsection{Fractals} While the KPZ dynamics is defined in an Euclidean space of dimension $d+1$, the growing surface shows fractal features observed in experiments on  SiO$_2$ films~\cite{Ojeda00} or in the rough interface generated by simulations of the $2+1$ single-step (SS)  model~\cite{Daryaei20}. The existence of an associated fractal dimension is widely known~\cite{Barabasi95,Kondev00}.

For these self-affine growth processes, the grown surface has a fractal dimension $d_f$, which obeys~\cite{Barabasi95}
\begin{equation}
\label{df}
d_f=
\begin{cases}
 2-\alpha , &\text{ if~~ }  d=1,2\\
 d-\alpha, &\text{ if~~ } d \geq 2.\\
\end{cases}
\end{equation}
Therefore, KPZ growth is a phenomenon intricately linked to fractality.
Moreover, dynamics of  complex systems like KPZ  can exhibit various length scales and, consequently, different fractal dimensions. With our current knowledge, we certainly cannot specify how many. Nevertheless, our primary focus here is to highlight two specific fractal dimensions: the previously mentioned $d_f$ and a new fractal dimension $d_n$ associated with the effective noise~$\phi$. 

{To motivate the need for a description in terms of a new fractal dimension, let us first recall that the system is defined in a space with dimension $d+1$, where ``1'' is associated with the height coordinate $h$. However, notice that the dynamical evolution of the KPZ equation leads to structures with effective dimension \emph{lower} than $d+1$ --- this becomes apparent in the long-time behavior associated with $w$, which scales as $L^\alpha$, with $\alpha<1$. Since this consideration only involves coordinate $h$, it is reasonable to consider an effective description in which the dynamics is embedded in a space with a putative lower dimension $d_n + 1$, so that $d \leq d_n + 1 \leq d +1$, i.e. $ d-1 \leq d_n \leq d $.}

{The argument above suggests the existence of a new fractal dimension, but it does not provide a workable definition for measuring or calculating $d_n$. One possibility to incorporate $d_n$ is partly motivated by recent results (see e.g.~\cite{Lima24}), and consists in replacing $d$-dimensional Dirac delta functions by $d_n$-dimensional \emph{fractional} delta functions~\cite{Jumarie09,Muslih10b}, which naturally incoporate non-locallity and correlations in space. Recall that our new noise variable $\phi$ must be correlated, so we make the simple conjecture that the two-point correlation function $\langle \phi(\vec{x},t) \phi(\vec{x'},t') \rangle$ can be written as
\begin{equation}
\label{FDT2}
\langle \phi(\vec{x},t) \phi(\vec{x'},t') \rangle  = 2D_{\rm eff}(t)\delta^{(d_n(t))}(\vec{x}-\vec{x'})\delta(t-t'),
\end{equation}
where both $d_n(t)$ and $D_\text{eff}(t)$ are functions of time, reflecting the fact that surface roughness evolves over time.
If we start with a flat interface, implying initial roughness $w(t=0)=0$, it will evolve until saturation at $t \gg t_\times$. Therefore, one has that $w(t) \rightarrow w_s$, $D_{\rm eff}(t)\rightarrow D_{\rm eff}^s$ and  $d_n(t) \rightarrow d_n^s$, where ``s'' indicates saturation values.
In Sec.~\ref{subsec:Dimensional}, we will use simple ideas based on dimensional analysis to connect the fractal dimension $d_n$ with the exponent $\alpha$.}

 Through this new perspective, there is actually no violation of the FDT: Eq.\ \eqref{FDT2} is understood as a real representation of fluctuations in the system. At saturation, the balance represented by the new FDT is an equilibrium between the dissipation of roughness $\nabla^2h$ and the fluctuation $\phi$. In Eq.\ (\ref{KPZm}), $v_g$ is a constant that does not contribute to this balance. We can now seek to associate $\alpha$ with $d_n$  for $d+1$ dimensions. 

\subsection{Dimensional analysis}
\label{subsec:Dimensional}
A powerful tool in physics is dimensional analysis, which we apply now to get important information about the interface geometry. 
{Although $w_s\sim L^\alpha$ as seen in Eq.\ (\ref{Sc1}), it has the same physical dimension as the height $h$, that is, 
$[w_s]=[h]=[L]$. In other words, in experiments they are both measured in units of length, as it must be from definition (\ref{W2}).
The physical dimensions involved in the parameters that control $w_s$ are $[\nu]=[L^2][T^{-1}]$, $[ D_{\rm eff}^s]=[L^{d_n+2}][T^{-1}]$, where $[T]$ is the time dimension.  Since time is not present in the dimensions of $w_s$, it needs to be eliminated. Therefore, both $ D_{\rm eff}^s$ and $\nu$ must appear under the same exponent in the form  $ D_{\rm eff}^s/\nu$.
Thus, the FDT balance gives  $w_s \propto ( D_{\rm eff}^sL/\nu)^\alpha$, whose dimensional analysis yields}
\begin{equation}
\label{alf}
\alpha=\frac{1}{d_n+1},
\end{equation}
with $d-1 \leq d_n \leq d$ as previously discussed.  For $d=1$, we have $d_n=d=1$. This is because if $d<1$, there would be no continuous border. Thus, for $1+1$ dimensions, our analysis yields the exact exponent $\alpha=1/2$.
}

\section{Determination of exponents and fractal dimensions}

Originally, there were three exponents and two equations, namely Eqs.\ (\ref{z}) and (\ref{GI}). We have now introduced Eqs.\ (\ref{df}) and (\ref{alf}). However, they involve two extra unknowns, $d_f$ and $d_n$, both associated with fractal dimensions. Although introducing these variables might seem pointless, it has the advantage of shifting our attention to the fractal geometry of the problem.

In the absence of a formal theory to determine at least one of the fractal dimensions, we will use computer simulations to obtain some information regarding the critical exponents. Knowing $\alpha$, we can then obtain the fractal dimensions $d_n$ and $d_f$ using the above relations. The surface roughness measured by the exponent $\alpha$ has important information on properties of the surface and of the growth process. Its evolution can be obtained from the correlation function:
\begin{equation}
\label{Corr}
C(r)= \left\langle [h(\vec{x}+\vec{r},t)- h(\vec{x},t)]^2\right\rangle \propto r^{2\alpha},
\end{equation}
where $r$ is the modulus of the vector $\vec{r}$ with $r < \xi$, where $\xi$ is the correlation length~\cite{Kondev00}.
Note that this can be viewed as a time-independent correlation function for each time $t$.

\begin{figure}[htbp]
\centering
\includegraphics[width=1\columnwidth]{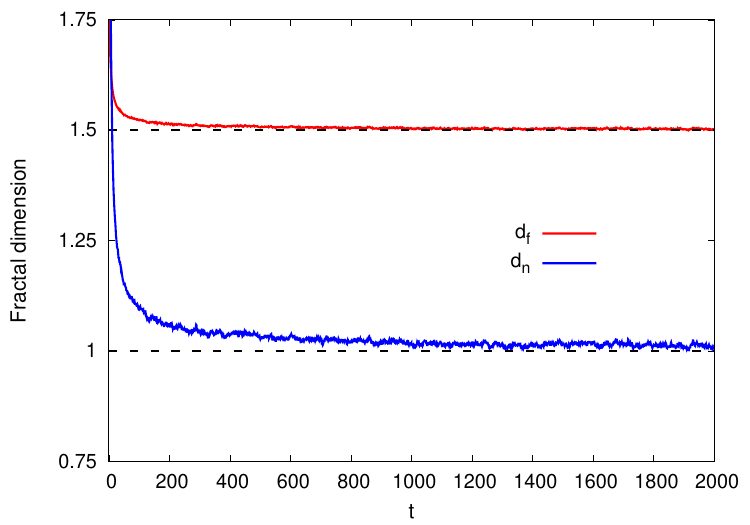}
\caption{Fractal dimensions $d_f$ and $d_n$ as a function of time $t$ for the SS model in $1+1$ dimensions. The dashed lines represent the stationary theoretical values for each fractal dimension (see text).}
\label{fig:1d}
\end{figure}

\begin{figure}[htbp]
\centering
\includegraphics[width=1\columnwidth]{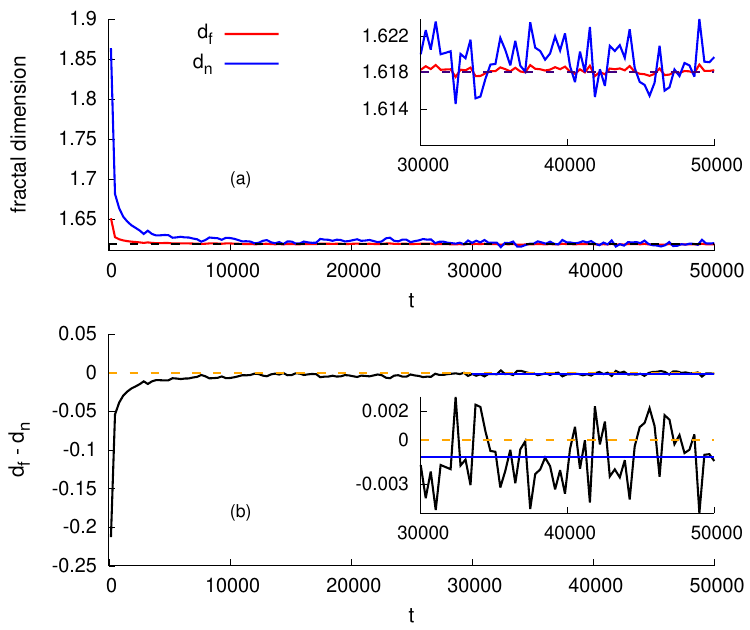}
\caption{SS model in $2+1$ dimensions. Top: Fractal dimensions $d_f$ and $d_n$ against time $t$. Dashed line represents the theoretical value for $d_n=d_f=\frac{1+\sqrt{5}}{2}$ (golden ration). Bottom: The difference between  the fractal dimensions,  $d_f - d_n$, as a function of time. {The dashed line marks zero, whereas the horizontal solid line represents the average value, $-0.0011(3)$, obtained within the time interval from $3\times10^4$ to $5\times10^4$.} 
In the insets, we zoom into the stationary regime data.}
\label{fig:2d}
\end{figure}

\begin{figure}[htbp]
\centering
\includegraphics[width=1\columnwidth]{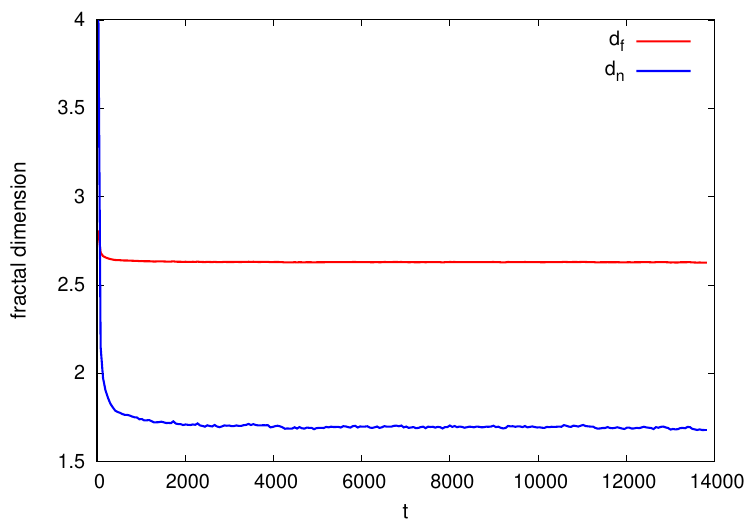}
 \caption{SS model in $3+1$ dimensions:  Fractal dimensions $d_f$ and $d_n$ as a function of time $t$.
} 
\label{fig:3d}
\end{figure}

Simulations using lattice models in the KPZ universaility class can be used to determine the time evolution of $\alpha(t)$, which in turn can be found by fitting the correlation function~\cite{GomesFilho21b,Luis22}. From that, we can obtain $d_n$ from  Eq.\ (\ref{alf}) and $d_f$ from Eq.\ (\ref{df}) as functions of time.
To achieve this, we simulate the well-known SS model as described below. The results are shown in Figures \ref{fig:1d} to \ref{fig:3d}.

 The SS lattice model is defined in such a way that the height difference between two neighboring heights, $\eta=h_i-h_j$, is always $\eta=\pm 1$.  Let us consider a hypercube of side $L$ and volume $V=L^d$. We will select a site $i$ and compare its height with that of its neighbors $j$, applying the following rules~\cite{Derrida98,Meakin86,Daryaei20}:
\begin{enumerate}
\item At time $t$, randomly choose a site $i\in{V}$;
\item If $h_i(t)$ is a local minimum, then $h_i(t+\Delta t)=h_i(t)+2$, with probability $p$;
\item If $h_i(t)$ is a local maximum, then $h_i(t+\Delta t)=h_i(t)-2$, with probability $q$.
\end{enumerate}

For all simulations presented here, we chose $p=1$ and $q=0$ to reduce computational time. Note that, if we implemented a simpler growth model based on rule $(1)$, one would have a white noise in $d+1$ dimensions. However, due to rules $(2)$ and $(3)$, only a fraction of that noise will be effectively realized. 

\begin{figure}[htbp]
\centering
\includegraphics[width=1\columnwidth]{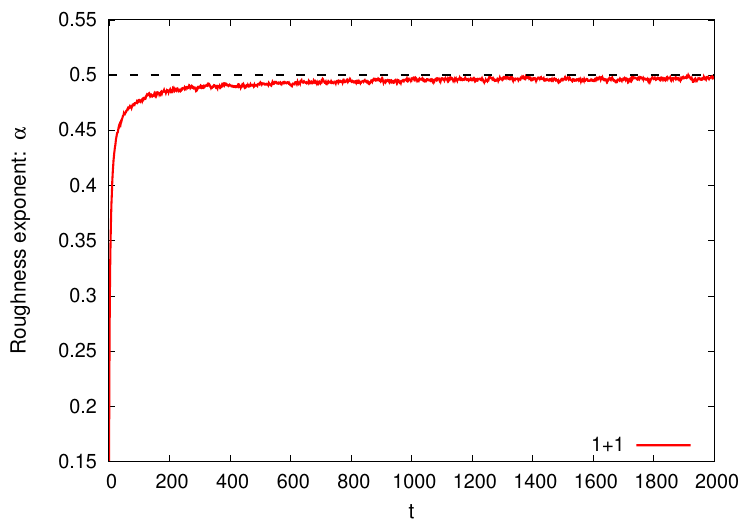}
\caption{ SS model in $1+1$ dimensions:  the roughness exponent $\alpha$ as a function of time t (in units of $t_\times$) for a  system of size  $L=4096$ obtained from the correlation function~(\ref{Corr}).   The dashed line represents the stationary theoretically-exact value for $\alpha$, i.e., 1/2.}
\label{fig:1d-alpha}
\end{figure}

We show in Figure~\ref{fig:1d-alpha} the time evolution of the roughness exponent $\alpha$ for the SS model in $1+1$ dimensions. The values are obtained from the correlation function~(\ref{Corr}) for a system of size $L=4096$. To do that, we average over the lattice [Eq.\ \eqref{W2}] and then over $1000$ experiments.
We observe that the value of $\alpha$ increases with time until it stabilizes, fluctuating around the stationary theoretically-exact value of $1/2$.

Having validated our simulations by comparison with the exact values, we now show in Figure \ref{fig:1d} the evolution of both fractal dimensions as functions of time for the SS model in $1+1$ dimensions, with  $d_f$  obtained from Eq.\ \eqref{df} and  $d_n$ from Eq.\ \eqref{alf}.
The simulation data are the same as used in Figure \ref{fig:1d-alpha}.  

We highlight that, since $\alpha$ increases over time and then saturates, the fractal dimensions $d_f$ and $d_n$ consequently decrease over time and then stabilize.
The stabilization occurs when the system reaches the saturation region where $w \approx w_s$. As $t \rightarrow \infty$, the value of $\alpha$ tends towards $1/2$. Consequently, $d_f \rightarrow 2 - \alpha = 3/2$ and $d_n \rightarrow 1/\alpha - 1 = 1$. These theoretical values are marked as dashed lines in Fig.~\ref{fig:1d}.

In Figure \ref{fig:2d} (top), we show the evolution of the fractal dimension as a function of time $t$ for the SS model in $2+1$ dimensions. The case of $2+1$ dimensions is  the most  relevant one. Besides corresponding to our real world, growth phenomena in these dimensions are associated with surface science and the development of new technological devices, such as those involving thin films. Moreover, for $2+1$ dimensions there are more simulation results available and one can get more precise exponents than, say, for $3 + 1$. Furthermore, for $2+1$ dimensions there are experimental results. We use a squared lattice of lateral size $L=2048$ and average over $10$ experiments. We also calculate the average over time windows of $500$ time steps. We determine $\alpha(t)$ and, from that, $d_f$ and $d_n$. 
Surprisingly, after the transient, the two values agree.  Figure \ref{fig:2d} (bottom) shows their difference $d_f-d_n$. In the inset, we see that, for a long time, the difference $d_f-d_n$ fluctuates around zero. Indeed, its mean value in the inset region is $\Delta d_f=\overline{d_f-d_n}=-0.0011(3)$. This yields $|\Delta d_f/d_f|= 7 \times 10^{-4}$.
Similar results, not presented here, hold for the etching model~\cite{Mello01,Rodrigues15,Alves16}.

Motivated by numerical evidence, we assume that $d_n=d_f$ for $2+1$ dimensions, which allows us to write down exact values for the exponents $\alpha$, $\beta$, $z$, as well as the fractal dimensions $d_f$ and $d_n$.
Combining Eqs.\ (\ref{df}) and (\ref{alf}), we obtain
\begin{equation}
\label{exp}
 \alpha=\frac{3-\sqrt{5}}{2}; \hspace{0.5cm}  \beta=\sqrt{5}-2;  \hspace{0.5cm}z =d_f=\frac{1+\sqrt{5}}{2},  
\end{equation}
which corresponds to $d_f=1.61803...$ (see inset of Figure \ref{fig:2d}, top), and $\alpha=0.381966011...$, in agreement with simulations (see compilations of simulation results in reference~\cite{GomesFilho21}). Moreover, accurate experiments give  $z=1.6(2)$~\cite{Orrillo17}, $z=1.6(1)$~\cite{Ojeda00},   $z=1.61(5)$~\cite{Almeida14}, and $z=1.61$ \cite{Fusco16} in agreement with our value of $z =d_f=\frac{1+\sqrt{5}}{2}=1.61803...$. Since the final fate of a theory is decided by experiments,  these results strongly indicate that our proposal is on the right track.
For completeness, we mention that, recently, Luis {\textit{et al}.}~\cite{Luis22,Luis23}  have used the Higuchi method (HM)~\cite{Higuchi88,Helmut15} and the {three-point sinuosity method} ~\cite{Zhou15}
to obtain $d_f = 1.6179(3)$ for the SS model and
$d_f=1.61813(5)$ for the etching model~\cite{Luis22} and discuss its theoretical and experimental accessibility during film growth~\cite{Luis23}.

For $3+1$  dimensions  the distinction between $d_n$ and $d_f$ becomes clear again. In Figure \ref{fig:3d} we use a cube of side $L=512$ and we average over 3 experiments and time windows of 50 time steps. The figure exhibits the evolution of both fractal dimensions. There is no doubt they correspond to different fractal dimensions.

\section{Additional discussion}
\subsection{Upper critical dimension}

For $d \geq 2$, no exact results for the KPZ exponents have been widely accepted. Equation~(\ref{alf}) may shed some light on the issue. From $d-1 \leq d_n \leq d$, we obtain:
\begin{equation}
\label{alf2}
\frac{1}{d} \geq \alpha \geq \frac{1}{d+1}.
\end{equation}
Therefore, $\alpha$ will keep changing  with the dimension $d$. As a consequence, within our framework, there is no upper critical dimension . Note that if we choose the bounds allowed by the Hausdorff fractal dimensions \cite{Barabasi95}, not the above restriction, we have $d-1 \leq d_n \leq d+1$, and therefore Eq.\ (\ref{alf}) implies
\begin{equation}
\label{alf3}
\frac{1}{d} \geq \alpha \geq \frac{1}{d+2}.
\end{equation}
Both sets of inequalities suggest the nonexistence of a UCD. However, $\alpha=(d+1)^{-1}$,  is the well-known Wolf-Kertesz relation~\cite{Wolf87}, which is broadly recognized as a lower bound for $\alpha$. Furthermore, the upper bound of $d-1 \leq d_n \leq d$ gives the exact result $\alpha=1/2$ for $d=1$ as already mentioned.  Thus, Eq.\ (\ref{alf2}) establishes the appropriate bounds and  we do not need relation (\ref{alf3}).

\subsection{Renormalization}

Equation (\ref{alf}) also sheds light on a crucial aspect of the one-loop renormalization approach~\cite{Kardar86}. For $d=1$, where the noise dimension $d_n=1$ aligns with the Euclidean dimension, this renormalization approach is correct. However, for $d \geq 2$, where $d_n$ differs signifcantly from $d$, it does not work. This mismatch between the two dimensions suggests an explanation as to why the one-loop renormalization approach is incorrect.

{The main relationships between exponents are the result of scaling, Eq.\ (\ref{z}), and renormalization approaches, Eq.\ (\ref{GI}). Recent results~\cite{Rodrigues24} generalizing the Family-Vicsek relation to all $d$ dimensions would be a hopeful starting point for a generalization of a renormalization group (RG) approach to KPZ.  Thus, a new approach involving a suitable renormalization with a fractal dimension for the noise would be desired. However, that is not an easy task.}

\subsection{A possible connection between growth and phase transitions}

We discuss above the violation and necessary modification of the FDT in growth. The first clear indication of FDT violation appeared in phase transition studies. For example, let us define the fluctuation of the order parameter $m(\vec{r},t)$ as $\delta m(\vec{r},t)= m(\vec{r},t)-\langle  m(\vec{r},t)\rangle$.
We define as well  the correlation function, $G(r)=\langle \delta m(\vec{r}+\vec{i},t)\delta m(\vec{i},t)\rangle$, which for small fluctuations in the continuous limit yields~\cite{Kardar07}
\begin{equation}
\label{G2}
G(r) \propto
\begin{cases}
r^{2-d} \exp(-r/\rho) , &\text{ if~~ } r>\rho,\\
r^{2-d-\eta}, &\text{ if~~ }  r \ll\rho,\\
\end{cases}
\end{equation}
 where $\rho $ is the correlation length.
At this point the Fisher exponent $\eta$ is introduced empirically, arguing that the FDT does not work.
Part of this is empirical, motivated by experiments and simulations.
But $\eta$ is also exactly calculated in some few exactly solvable models (e.g. $\eta = 1/4$ for Ising in 2D).
A recent fractal  analysis~\cite{Lima24} close to the phase transition shows that $G(r)$ is the appropriate  response function with
\begin{equation}
\label{dff}
 \eta=d-d_f.
\end{equation}
Thus, the Fisher exponent in the correlation function,  $G(r)$, represents the deviation from the integer dimension. Note the similarity with Eq.\ (\ref{df}). Such similarity is remarkable since we are comparing non-equilibrium growth phenomena with equilibrium phase transitions.

\section{Conclusion}

 In this work, our objective was to  give a new insight into the fluctuation dissipation theorem for the KPZ equation. To do this, we consider the fluctuation of a combination of the nonlinear term with the white noise. Our theory suggests a new emergent noise which obeys a new FDT with fractal dimension $d_n$. The balance at saturation $w \approx w_s$ gives a new equation relating $d_n$ to the exponent $\alpha$. 
 This new relation indicates when one-loop RG should work or not. 
 For  $2 + 1$ dimensions the noise dimension and the fractal dimensions are the same within a great precision, $d_n \approx d_f$, which allows us to  obtain  accurate  values of the growth exponents in 2 + 1 dimensions for the KPZ equation.
 
 Finally, the discussions presented here open a new scenario for further investigation of different forms of growth both theoretical and numerical.  For example, the RG approach applied to the fractal interface will probably lead to new important results. As mentioned above, one-loop expansion preserves the Galilean invariance (\ref{GI}). However, it deserves further developments. The  attempt to obtain exact height fluctuations for the stationary KPZ equations, as well as for most of KPZ growth physics in  $2+1$ dimensions,  is still in its  beginning.  These theoretical methods will benefit  from the fixed points obtained by precise KPZ exponents, and from the idea of a fractal geometry that must be associated with them~\cite{Anjos21}. We expect as well that new methods would confirm our results. Therefore, our work suggests new horizons for KPZ research.

\section*{Acknowledgments}

This work was supported by the Funda\c{c}\~ao de Apoio a Pesquisa do Rio de Janeiro (FAPERJ), Grant No. E-26/203953/2022 (F.A.O.). M.S.G.F acknowledges financial support from grant number \# 2023/03658-9, São Paulo Research Foundation (FAPESP) and acknowledge the National Laboratory for
Scientific Computing (LNCC/MCTI, Brazil) for providing computing resources trough the SDumont
supercomputer.
{D.B.L. thanks financial support through FAPESP grants \# 2021/14285-3 and \# 2022/09615-7.} P.d.C. was supported by Scholarships \# 2021/10139-2 and \# 2022/13872-5 and ICTP-SAIFR Grant \# 2021/14335-0, all granted by São Paulo Research Foundation (FAPESP), Brazil.


\bibliography{references}

\end{document}